\documentclass[a4paper,11pt,twoside]{article}

\usepackage{epsfig}
\usepackage{amsmath}
\usepackage{amssymb}

\usepackage[abbr]{harvard}
\citationstyle{dcu}
\citationmode{abbr}
\renewcommand{\harvardand}{\&}

\newcommand{\DEF}{\stackrel{\mbox{\rm\scriptsize def}}{=}}

\begin{document}                                   %

\title{Large geometric phases and non-elementary monopoles} 
\author{ P. Leboeuf $^1$ and A. Mouchet $^2$\thanks{mouchet@celfi.phys.univ-tours.fr}} 
\date{$^1$Laboratoire de Physique Th\'eorique et Mod\`eles 
Statistiques\footnote{Unit\'e de recherche de l'Universit\'e de Paris XI
associ\'ee au CNRS} ,
B\^at. 100, \\ Universit\'e de Paris-Sud, 91405 Orsay Cedex, France. \\ 
$^2$Laboratoire de Math\'ematiques et de Physique Th\'eorique \textsc{(cnrs umr 6083)},
Universit\'e Fran\c{c}ois Rabelais, ave\-nue Monge, Parc de Grandmont, 37200
Tours, France.\\ }


\maketitle

\begin{abstract}
Degeneracies in the spectrum of an adiabatically transported quantum system
are important to determine the geometrical phase factor, and may be
interpreted as magnetic monopoles. We investigate the mechanism by which
constraints acting on the system, related to local symmetries, can create
arbitrarily large monopole charges. These charges are associated with
different geometries of the degeneracy. An explicit method to compute the
charge as well as several illustrating examples are given.
\end{abstract}


\newpage


\section{Introduction}
\label{sec:introduction}
When a quantum system is adiabatically transported along a closed
loop~$\mathcal{L}$ in a parameter space, in addition to the usual dynamical
phase every non-degenerate eigenstate accumulates an extra phase~$\Delta\Phi$.
For a 3\textsc{d}-parameter space, M. V. Berry gives a geometrical
interpretation of~$\Delta\Phi$ as the flux of a magnetic-like field
$\mathbf{B}$ through a surface with boundary $\mathcal{L}$ \cite{Berry84a}.
The degeneracies of the spectrum in parameter space are the singularities of
the vector field $\mathbf{B}$, and therefore play an important role in
connexion with the geometric phase. Each degeneracy can be seen as a charge
distribution located at the contact point between energy surfaces. Because the
eigenstates are smooth and single valued outside the degeneracies, the total
charge of the distribution, i.e. the monopole charge or moment, is necessarily
an integer multiple of the elementary charge $g_0=1/2$. In the generic case of
a diabolical contact \cite{Berry84a}, the monopole charges are precisely~$\pm
g_0$. However, higher integer multiples of~$g_0$ may occur. For instance, for
light propagating through a twisted anisotropic dielectric medium there are
experimental situations \cite{Berry86a} where the monopole charges
are~$\pm2g_0$. Other examples arise in condensed matter physics. In a
bidimensional periodic crystalline lattice, the Hall conductance of a gas of
independent electrons is proportional to a topological index, the Chern index,
that measures the net charge inside a closed surface associated to the first
Brillouin zone \cite{Simon83a}. In some models, large jumps of the Chern index
that cannot be explained by elementary charges have been
observed~\cite{Leboeuf+90a}~\cite{Faure/Leboeuf93a}.

Our purpose is to discuss a generic mechanism for the production of monopole
charges larger than $g_0$. The mechanism is due to constraints that act on the
system, and may be associated to local symmetries. Section~\ref{sec:context}
introduces the notation and some general formulae for~$\Delta\Phi$. The
generic case of a diabolical contact is briefly recalled in
section~\ref{sec:genericcase}. In section~\ref{sec:constraints}, which is the
central part of the paper, a broader class of situations that incorporates
constraints is considered. The consequences on the geometry of the contact
point and on the corresponding monopole charges are discussed, and a method to
explicitly compute them is introduced. The latter provides an alternative and
simple view of the quantization of the monopole charge as a sum over winding
numbers associated to Dirac strings. We illustrate the general approach by two
examples in section~\ref{sec:2examples}, and show how arbitrarily large
multiples of~$g_0$ can actually be generated in~\ref{sec:largecharges}. The
last section contains a discussion of some of the results obtained.

\section{Background material} 
\label{sec:context}
We consider a quantum system governed by a Hamiltonian depending on three real
parameters~$\mathbf{r}=(x,y,z)$. We denote~$\mathcal{P}$ the parameter space
and suppose that in the neighborhood of $\mathbf{r}=\mathbf{0}$ two
eigenvalues are close to each other. Hence, the coupling to other states can
be neglected and the Hamiltonian can be restricted to a bidimensional
eigenspace. The most general form of the reduced Hamiltonian is
\begin{equation}\label{def:H} H(\mathbf{r})\ =\
\begin{pmatrix} s(\mathbf{r})+e_z(\mathbf{r}) &
e_x(\mathbf{r})-\mathrm{i}\,e_y(\mathbf{r}) \\[3ex]
e_x(\mathbf{r})+\mathrm{i}\, e_y(\mathbf{r}) & s(\mathbf{r})-e_z(\mathbf{r})
\end{pmatrix} \ =\ s(\mathbf{r})+\text{\bfseries\itshape
e}(\mathbf{r}).\boldsymbol{\sigma}
\end{equation} 
$s(\mathbf{r})$ and the three components of $\text{\bfseries\itshape
e}(\mathbf{r})=[e_x(\mathbf{r}),e_y(\mathbf{r}),e_z(\mathbf{r})]$ are smooth
real functions in~$\mathcal{P}$.
$\boldsymbol{\sigma}=[\sigma_x,\sigma_y,\sigma_z]$ are the Pauli matrices.
 
The eigenvalues of $H(\mathbf{r})$ are given by
\begin{equation}\label{def:energies} 
	E_\pm(\mathbf{r})\ =\ s(\mathbf{r})\pm e(\mathbf{r}) 
\end{equation}
where~$e=\sqrt{e_x^2+e_y^2+e_z^2}$. The points~$\mathbf{r}$ of the parameter
space where degeneracies occur are given by the set~$\mathcal{M}$ where $e$
vanishes.

One possible orthonormal eigenbasis corresponding to~$E_\pm(\mathbf{r})$ is
\begin{equation}\label{def:vectors} |\psi_\pm(\mathbf{r})\rangle\ \DEF\
\frac{1} {\sqrt{2e(\mathbf{r})\, \bigl(e(\mathbf{r})\mp e_z(\mathbf{r})\bigr)
} } \begin{pmatrix} e_x(\mathbf{r})-\mathrm{i}\, e_y(\mathbf{r})\\[1ex] \pm
e(\mathbf{r})-e_z(\mathbf{r}) \end{pmatrix}\;. 
\end{equation} 
It follows from this expression that the eigenstates of~$H$ are not only
singular on~$\mathcal{M}$ but more generally on a larger subset~$\mathcal{D}$
of~$\mathcal{P}$. With the choice~(\ref{def:vectors}), $\mathcal{D}$ is given
by the points where~$e(\mathbf{r})\mp e_z(\mathbf{r})$ vanishes. For
convenience, we split $\mathcal{D}$ into two sets, $\mathcal{D}_+$
and~$\mathcal{D}_-$, that corresponds to $e_x(\mathbf{r})=e_y(\mathbf{r})=0$
and $e_z(\mathbf{r})\ge0$ ($e_z(\mathbf{r})\le0$), respectively. These two
sets intersect on~$\mathcal{M}$.

Let~$\mathcal{L}$ be any closed loop in~$\mathcal{P}$ not
intersecting~$\mathcal{D}$. Berry \cite{Berry84a} gave a {\em geometrical}
interpretation (that is, coordinate free in~$\mathcal{P}$) of the
non-dynamical part~$\Delta\Phi_\pm$ of the phase shift that
$|\psi_\pm(\mathbf{r})\rangle$ acquires when~$H(\mathbf{r})$
follows~$\mathcal{L}$ adiabatically. Simon~\cite{Simon83a} completed the
picture with a {\em topological} interpretation of Berry's phase in terms of
connection on a suitable fiber bundle. Topological arguments were anticipated
in~\cite[see in particular the section:``A topological test for
intersections'', p.85]{Stone76a}. The phase shift after one traversal
of~$\mathcal{L}$ is given by the circulation
\begin{equation}\label{eq:circulation}
	\Delta\Phi_+=-\Delta\Phi_-=\Delta\Phi\ \DEF\ 
	\oint_\mathcal{L}\mathbf{A}(\mathbf{r}).\,d\mathbf{l}
\end{equation}
of the vector $\mathbf{A}\DEF-\mathrm{Im}\Bigl[
\langle\psi_+|\boldsymbol{\nabla_{\!r}}|\psi_+\rangle\Bigr].$ By Stokes's
theorem, $\Delta\Phi$ can alternatively be computed from,
\begin{equation}\label{eq:flux}
	\Delta\Phi=\int_\mathcal{S}\mathbf{B}(\mathbf{r}).\,d\boldsymbol{\Sigma}\;.
\end{equation}
$\mathcal{S}$ is any surface in~$\mathcal{P}$ not intersecting~$\mathcal{D}$
whose border is~$\mathcal{L}$. $\mathbf{B}(\mathbf{r})$ is the \hbox{3-vector}
field (more generally a \hbox{2-form}) obtained by taking the curl
of~$\mathbf{A}$. Specifically, outside~$\mathcal{D}$ and in the
two--dimensional subspace where our analysis is restricted
\begin{equation}
	\mathbf{B}= \frac{ \langle\psi_+| \boldsymbol{\nabla_{\!r}}\,H
 	|\psi_-\rangle \wedge \langle\psi_-| \boldsymbol{\nabla_{\!r}}\,H
 	|\psi_+\rangle } {(E_+-E_-)^2}\;.
\end{equation} 
From~(\ref{def:H}), (\ref{def:energies}) and~(\ref{def:vectors}) we may write,
outside~$\mathcal{D}$,
\begin{equation}\label{eq:A}
	 \mathbf{A}
         =\frac{e_y\boldsymbol{\nabla_{\!r}}\,e_x-e_x\boldsymbol{\nabla_{\!r}}\, e_y}
               {2(e_x^2+e_y^2)}
	\left(1+\frac{e_z}{e}\right)\;, 
\end{equation} and
\begin{equation}\label{eq:B}
	\mathbf{B}
	=\frac{1}{2e^3} \Bigl[
	e_x(\boldsymbol{\nabla_{\!r}}\, e_y\wedge\boldsymbol{\nabla_{\!r}}\, e_z)
      + e_y(\boldsymbol{\nabla_{\!r}}\, e_z\wedge\boldsymbol{\nabla_{\!r}}\, e_x)
      + e_z(\boldsymbol{\nabla_{\!r}}\, e_x\wedge\boldsymbol{\nabla_{\!r}}\, e_y) 
                            \Bigr]\;.
\end{equation} 

\section{Generic case (no constraints)}\label{sec:genericcase}

Generically~$\mathcal{M}$ is either empty or an isolated point
(see~\cite{vonNeumann/Wigner29a}). We consider the latter case and work in a
local chart centered at the contact point, that we assume is located
at~$\mathbf{r}=\mathbf{0}$. Since generically
\begin{equation}\label{def:D}
	\Lambda\ \DEF\ \det\bigl[ \boldsymbol{\nabla_{\!r}}\, e_x,
\boldsymbol{\nabla_{\!r}}\, e_y, \boldsymbol{\nabla_{\!r}}\, e_z
\bigr]_{\left|\mathbf{r}=\mathbf{0}\right.}  
\end{equation} 
is different from zero, by virtue of the local inversion theorem we can
consider~$(e_x,e_y,e_z)$ to be a new local chart in a neighborhood
of~$\mathbf{r}=\mathbf{0}$. In other words, we can
take~$\mathbf{r}=\text{\bfseries\itshape e}$. This is the generic case. The
two energy surfaces~$E_\pm(\mathbf{r})$ intersect through an isolated
doubly-conical contact point (diabolical point) \cite{vonNeumann/Wigner29a}.
From~(\ref{eq:B}) we have
\begin{equation}\label{eq:elementaryB}
	\mathbf{B}(\mathbf{r})=\frac{\mathbf{r}}{2||\mathbf{r}||^3} \ .
\end{equation} 
Close to a conical intersection the vector~$\mathbf{B}$ corresponds therefore
to the field created by a magnetic monopole of
charge~$g=g_0=1/2$~\cite{Dirac31a,Dirac78a} (for the
state~$|\psi_-(\mathbf{r})\rangle$ the opposite charge must be taken).
$\mathbf{A}(\mathbf{r})$ is the vector potential associated to it. The
sets~$\mathcal{D}_\pm$ are the well known Dirac (half) strings which
physically can be interpreted as semi-infinite solenoids (with end points
at~$\mathbf{r}=\mathbf{0}$) carrying a magnetic flux~$\pm 4\pi g_0$,
respectively. This magnetic analogy cannot however be extended globally since
there is no superposition theorem available here.

\section{Finite number of constraints}
\label{sec:constraints}

\subsection{Non-elementary monopoles} 

\begin{figure}[!hbt]
\begin{center}
\psfig{file=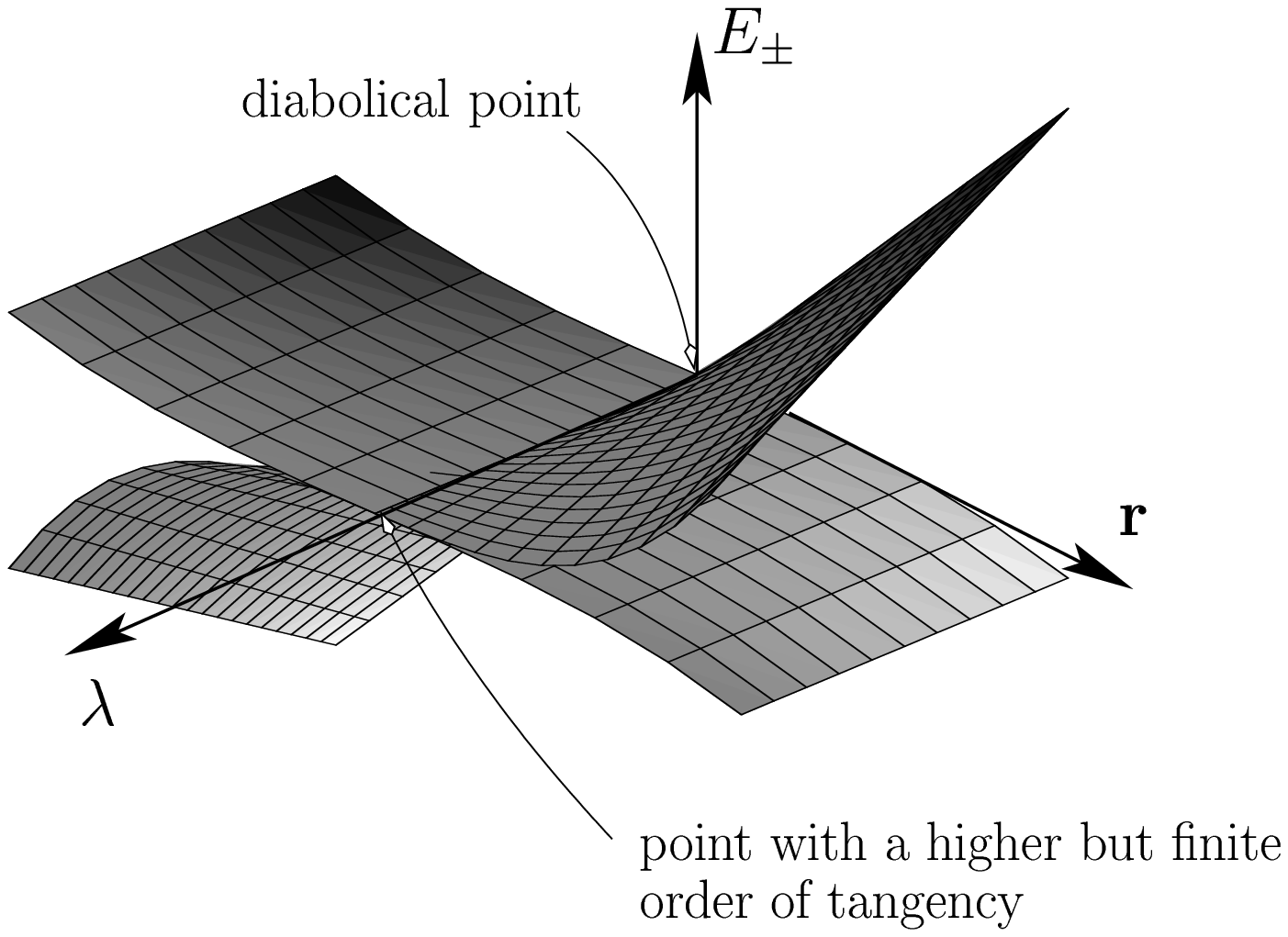,width=\textwidth}
\caption{The unfolding shown in this picture illustrates the changes in the
geometry of the contact point between two energy surfaces induced by the
variation of some control parameter~$\lambda$.}
\label{fig:Econstraints}
\end{center}
\end{figure}

We now consider situations where some constraints acting on the system are
present. These constraints can be thought of as independent real control
parameters~$\lambda=\{\lambda_n\}_{n\in\{0,\ldots,C-1\}}$ that impose~$C$
independent relations between the derivatives of~$\text{\bfseries\itshape e}$
at~$\mathbf{r}=\mathbf{0}$. Physically the control parameters are distinct
from the adiabatic parameters~$\mathbf{r}$ in that they reflect a local
symmetry in~$\mathcal{P}$ and therefore do not change during the adiabatic
transport along~$\mathcal{L}$. Due to these additional constraints, it may
happen that the determinant~$\Lambda$ defined by Eq.~(\ref{def:D}) vanishes.
The latter condition requires generically the existence of one external
parameter~$\lambda$ in addition to~$\mathbf{r}$. In that case, the contact
between the energy surfaces \emph{restricted to
the~$\mathbf{r}$-space~$\mathcal{P}$} is not a diabolical point anymore (see
figure~\ref{fig:Econstraints}). For instance, the charge $g= 2 g_0$
occasionally observed in Ref.~\cite{Leboeuf+90a} was found to correspond to a
parabolic contact. By increasing the number of parameters $\lambda$ one can
accordingly cancel an increasing number of derivatives
of~$\text{\bfseries\itshape e}$ at~$\mathbf{r}=\mathbf{0}$, and therefore
increase accordingly the degree of tangency of the surfaces defined
by~(\ref{def:energies}).

If the number of constraints~$C$ is finite, in the full $(C+3)$--dimensional
parameter space (that includes both the adiabatic and the control parameters),
the set of degeneracies is~$C$-dimensional and generically crosses
transversely at an isolated point located at~$\mathbf{r}=\mathbf{0}$ any
3-dimensional space defined by~$\lambda=\mathrm{constant}$. It requires an
infinite number of constraints to transform~$\mathcal{M}$ into a continuous
family of points since an infinite number of vanishing derivatives
of~$\text{\bfseries\itshape e}$ is needed. This is the situation when, for
instance, the system is invariant under a \emph{global} symmetry, like the
time reversal symmetry. In that case~$e_y\equiv0$ at any point
in~$\mathcal{P}$ and~$\mathcal{M}$ is a 1-dimensional sub-manifold
of~$\mathcal{P}$. Global symmetries that increase the dimension of the contact
manifold are not considered here, we restrict to the cases where the number of
constraints due to local symmetries is finite.

Under these assumptions, it follows from~(\ref{eq:B}) that $\mathrm{div}\,
\mathbf{B}$ is a distribution with support in $\mathbf{r}=\mathbf{0}$. It can
therefore be expanded as,
\begin{equation}\label{eq:divexpansion}
\mathrm{div}\, \mathbf{B} =
g\, \delta(\mathbf{r}) + \text{terms involving derivatives of~$\delta(\mathbf{r})$.}
\end{equation}
As for an arbitrary circuit in parameter space \cite[p. 55]{Berry84a}, and in
contrast to the diabolical contact, generically in the two--dimensional
subspace of the degeneracy the geometric phase is not completely determined by
\eqref{eq:divexpansion} because of the non-vanishing curl of~\eqref{eq:B}. Our
purpose is to determine an explicit rule to compute~$g$. A more detailed
qualitative discussion of the contribution to the geometric phase of the curl
of $\mathbf{B}$ and of the additional multipolar terms
in~\eqref{eq:divexpansion} will be made in the final section.

\subsection{The monopole charge as a sum of winding numbers}

One of the most remarkable properties of the monopoles is the quantization of
their charge, which can only take integer multiple values of~$g_0=1/2$. This
easily follows from a topological argument given by \cite{Stone76a}. Consider
a small loop~$\mathcal{L}$ that is smoothly retracted to a point without
crossing~$\mathcal{D}$. Because the eigenstates remain smooth and single
valued, $\Delta\Phi$ must tend to an integer multiple of~$2\pi$. But
$\Delta\Phi$ is the flux of~$\mathbf{B}$ through any surface~$\mathcal{S}$
with boundary~$\mathcal{L}$. When~$\mathcal{L}$ shrinks to a point,
$\mathcal{S}$ tends either to a point or to a closed finite surface that
possibly encloses a monopole charge~$g$. Therefore~$\Delta\Phi\to4\pi g$,
which must be an integer multiple of~$2\pi$. The quantization of~$g$ follows.

This argument is, however, unable to provide a method to compute the charge.
One usually has to explicitly compute the integral (\ref{eq:flux}) (using
(\ref{eq:B}) for the magnetic field) over a sphere enclosing the degeneracy.
In the case of a conical contact, the structure of the integral is
transparent: $\mathbf{B}.\,d\boldsymbol{\Sigma}$ is half the solid angle
subtended from the origin by $d\boldsymbol{\Sigma}$. However, for an arbitrary
contact point in the presence of additional constraints
$\mathbf{B}.\,d\boldsymbol{\Sigma}$ is a complicated function of the polar and
azimuthal angles. The structure of the two-dimensional integral is therefore
non-trivial and in general difficult to compute. It obscures moreover the
simplicity of the result (an integer multiple of $2 \pi$).

Rather than the magnetic field~$\mathbf{B}$, from now on we work with the
potential~$\mathbf{A}$. As we will show, the 2D-flux integral across the
sphere is replaced by one or several 1D-circulation integrals. This scheme
leads to a general and simple rule for calculating the charge that explicitly
exhibits the quantization.

When the number of constraints~$C$ is finite, the subset in~$\mathcal{P}$
defined by~$e_x(\mathbf{r})=e_y(\mathbf{r})=0$ is the union of the algebraic
curves defined near~$\mathbf{r}=\mathbf{0}$ by the first non-vanishing terms
of the Taylor expansion of {\bfseries\itshape e}. The expected ``pathologies''
of this subset are intersections of the curves at~$\mathbf{r}=\mathbf{0}$.
When the conditions~$e_z(\mathbf{r})\ge0$ and~$e_z(\mathbf{r})\le0$ -- that
define the Dirac strings~$\mathcal{D}_+$ and~$\mathcal{D}_-$ -- are added, the
number of curves can possibly be reduced by half. In a small enough
neighborhood of~$\mathbf{r}=\mathbf{0}$, ~$\mathcal{D}_+$ and~$\mathcal{D}_-$
are made of sets of half curves starting at the origin. Locally one cannot
determine if a given couple of half strings belongs to the same algebraic
curve or not. We also stress that it is the choice of gauge which
fixes~$\mathcal{D}_+$. Adding a gradient to~$\mathbf{A}$ will not change the
value of~$\Delta\Phi$ but can modify~$\mathcal{D}_+$ significantly by changing
the position and the number of half strings. What really matters is the
algebraic sum of the flux they carry. To illustrate this point, a useful gauge
transformation is to make a rotation~$\mathsf{R}\in\mathrm{SO}(3)$
on~$\text{\bfseries\itshape e}$: If~$U$ denotes a unitary representation
of~SO(3) in the bidimensional Hilbert subspace, the relation
\begin{equation}
\bigl[U(\mathsf{R})\bigr]^\dagger\,\text{\bfseries\itshape
e}\,.\,\boldsymbol{\sigma}\,U(\mathsf{R}) =\mathsf{R}(\text{\bfseries\itshape
e})\,.\,\boldsymbol{\sigma}\;, 
\end{equation} 
together with~(\ref{def:H}), implies that such a rotation corresponds to a
unitary transform of the eigenvectors of~$H$ and hence will not affect the
phases. Yet a simple permutation of the components of~$\text{\bfseries\itshape
e}$ can change~$\mathcal{D}_+$ drastically. For example, a rotation by~$\pi$
about~$e_x$ changes the sign of~$e_z$ without modifying~$e_x^2$ and~$e_y^2$.
One can therefore exchange~$\mathcal{D}_+$ and~$\mathcal{D}_-$. It is moreover
not excluded that a gauge transformation that considerably simplifies the
sets~$\mathcal{D}_+$ and~$\mathcal{D}_-$ exists in general, by reducing them
for example to a single half Dirac string. But no general rule for
constructing such a gauge transformation is available.

\begin{figure}[!ht]
\begin{center} 
\psfig{file=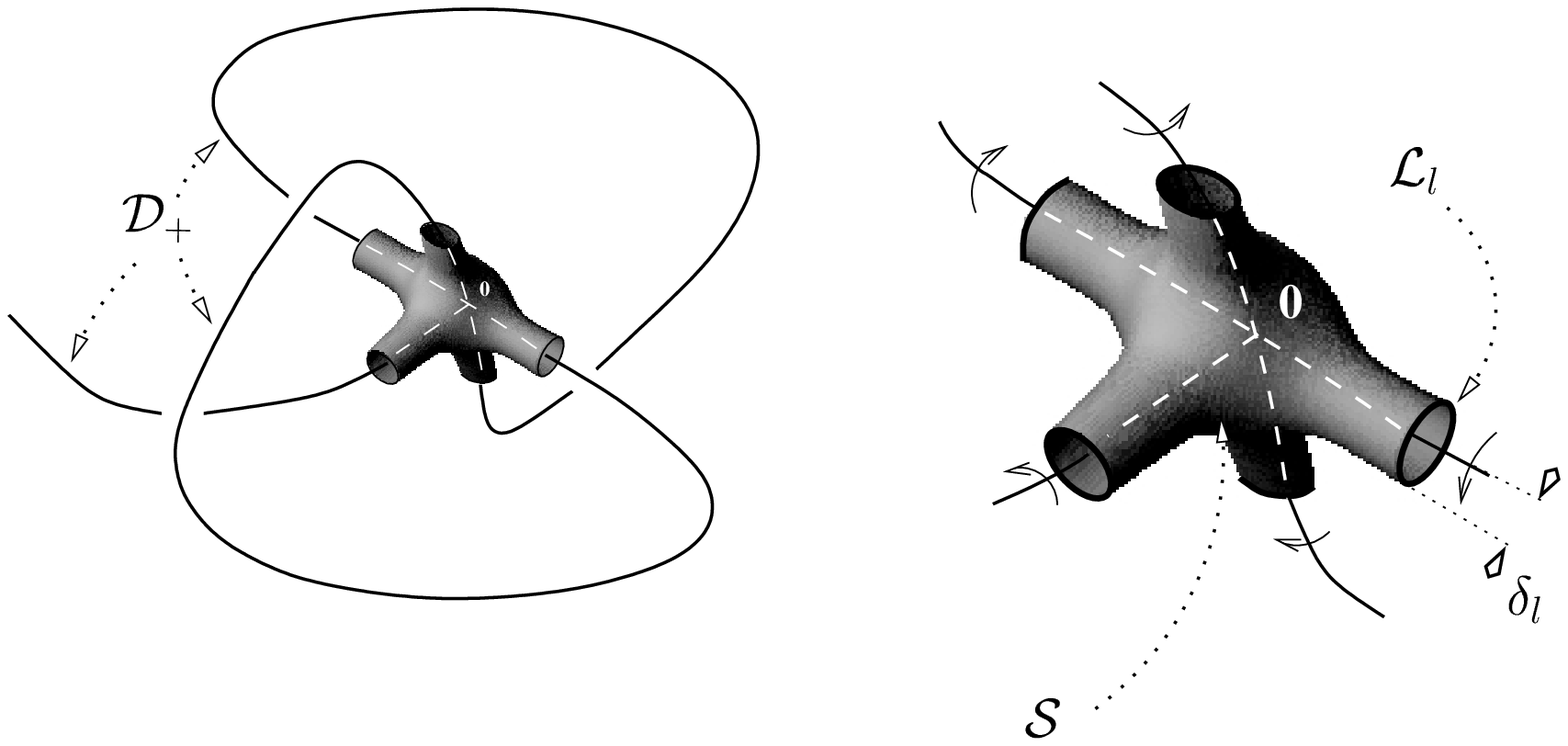,width=\textwidth}
\caption{Example of a set~$\mathcal{D}_+$ in the presence of constraints. On the
left part of the figure is displayed a closed Dirac string crossing itself
once at~$\mathbf{r}=\mathbf{0}$, in addition to a half Dirac string. From a
global perspective, one can predict that the former will not contribute to the
monopole charge. Nevertheless, in a neighborhood of~$\mathbf{r}=\mathbf{0}$
(right figure) one has to compute the winding of~$\mathcal{Z}_l$
when~$\mathbf{r}$ traverses~$\mathcal{L}_l$ for each of the five
half-strings.
\label{fig:diracstrings} }
\end{center} 
\end{figure}

From now on we fix the gauge according to Eq.~(\ref{eq:A}). $\mathcal{S}$ is
chosen to be a surface which is diffeomorphic to a sphere centered
at~$\mathbf{r}=\mathbf{0}$ but pierced with~$N$ holes at the level of the
strings contained in~$\mathcal{D}_+$. We denote~$\mathcal{L}_l$ the (oriented)
boundary of the~$l$th hole. It is a small loop of typical radius~$\delta_l>0$
that encircles the~$l$th half string. Then the integration contour
in~(\ref{eq:circulation}) reduces simply
to~$\partial\mathcal{S}=\mathcal{L}=\sum_{l=1}^N\mathcal{L}_l$ (see
figure~\ref{fig:diracstrings}).

Since $\mathcal{L}$ can be taken arbitrarily close to~$\mathcal{D}_+$, and not
intersecting~$\mathcal{D}_-$, we have
\begin{equation}
  \frac{e_z(\mathbf{r})}{e(\mathbf{r})}=1+O(\delta_l)\quad\mathrm{when}\quad
         \mathbf{r}\in\mathcal{L}_l\;.
\end{equation}
From this it follows that
\begin{equation}\label{g}
         g=\lim_{\mathcal{S}\to \text{closed sphere}} \frac{\Delta\Phi}{4\pi}\
         = \frac{1}{2} \sum_{l=1}^N w_l\;,                   
\end{equation}
where
\begin{equation}
	w_l\DEF \frac{1}{2\pi}\ \oint_{\mathcal{L}_l}\;
  \frac{e_y\boldsymbol{\nabla_{\!r}}\,e_x-e_x\boldsymbol{\nabla_{\!r}}\,e_y}{e_x^2+e_y^2}
                     \;d\mathbf{l}_l\;.
\end{equation}
$w_l$ has a natural topological interpretation. It is the algebraic winding
number of the complex number
\begin{equation}\label{def:ez}
         \mathsf{Z}_l(\mathbf{r})\DEF e_x(\mathbf{r})+\mathrm{i}\,e_y(\mathbf{r})
\end{equation}
when~$\mathbf{r}$ traverses~$\mathcal{L}_l$.
It is a well defined quantity: $\mathsf{Z}_l(\mathbf{r})$ never crosses zero
because~$\mathbf{r}$ never touches~$\mathcal{D}$.

Eq.~(\ref{g}) yields an alternative interpretation of the charge as a sum of
winding numbers associated to the Dirac strings. It also provides a method to
compute the charge, or to generate arbitrary values.

\section{Two examples}\label{sec:2examples}

We end up by illustrating the procedure with several examples (the generic
case of section~\ref{sec:genericcase} is similar to the second one). In all
the cases described below, we will Taylor expand~$\text{\bfseries\itshape e}$
near the origin and suppose that the terms that we explicitly retain are
sufficient to fix the local geometry of~$\mathcal{D}$ completely.

\subsection{An example with two choices of gauge}
\label{subsec:4diracstrings}
\begin{figure}[!ht]
\begin{center}
\psfig{file=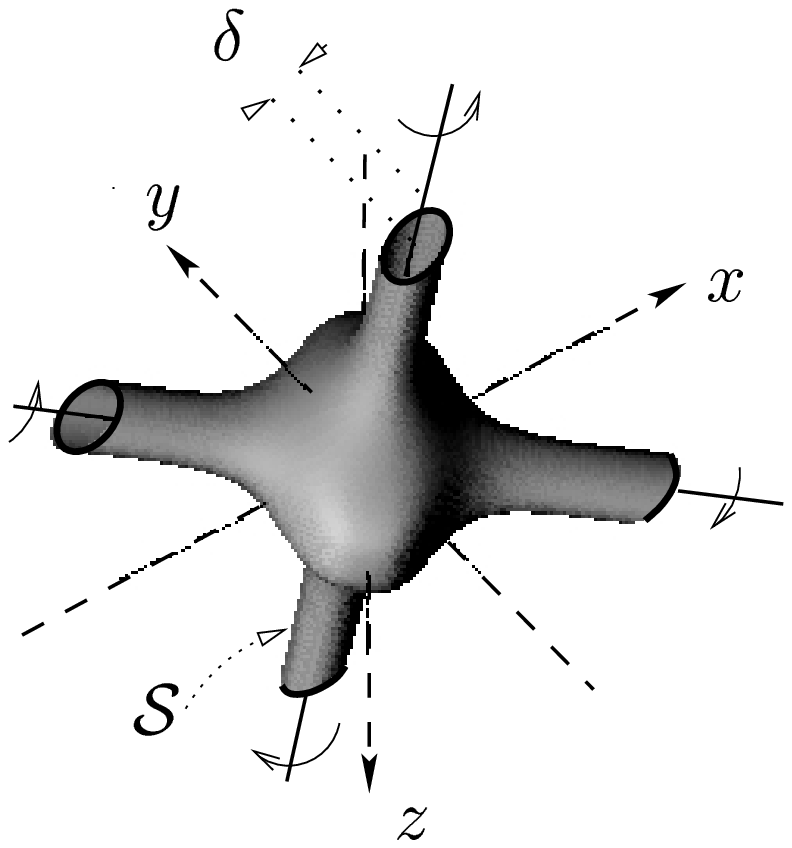}
\end{center}
\caption{\label{fig:example} 
Near~$\mathbf{r}=\mathbf{0}$, the Dirac strings in~$\mathcal{D}_-$ are tangent to
the two straight lines~$x=\pm y$, and the four loops~$\mathcal{L}_l$ are
circles of radius~$\delta$ and constant~$y$ with center in one of the lines.}
\end{figure}

Consider a case where in a suitable choice of coordinates we have,
\begin{equation}\label{def:be1}
\text{\bfseries\itshape e}(x,y,z)\ =\ [ x^2+z,y^2+z,z ]+\text{higher order terms}\;.
\end{equation}
It follows immediately that~$\mathcal{D}_+=\{\mathbf{0}\}$ so that
$g=0$. To illustrate the importance of the gauge, we now treat the same
example but in a gauge that exchanges~$\mathcal{D}_+$ and~$\mathcal{D}_-$.
$\mathcal{D}_-$ is made of four half branches of parabolae:
$\bigl\{[x=\pm\sqrt{-z}, y=\pm\sqrt{-z}, z\le0]\bigr\}$. Close enough
to~$\mathbf{r}=\mathbf{0}$ they can be assimilated to their tangent, i.e. the
two axes $\{x=\pm y,z=0\}$. The surface~$\mathcal{S}$ looks topologically like
the one in figure~\ref{fig:example}.

The four small loops~$\mathcal{L}_l$ can be chosen to be centered
on~$\mathcal{D}_-$ and parametrized by $[x=\pm \epsilon+\delta\cos\theta,
y=\pm \epsilon, z=-\epsilon^2-\mathrm{sgn}(y)\delta\sin\theta]$. $\theta$ runs
from $0$ to $2\pi$, whereas $\epsilon$ and~$\delta$ are fixed and small
strictly positive quantities measuring the distance to the origin and the
radius of the loops, respectively. From the definitions~(\ref{def:ez})
and~(\ref{def:be1}) we get $\mathsf{Z} =
\delta[-\mathrm{sgn}(y)(1+\mathrm{i})\sin\theta + 2 \
\mathrm{sgn}(x)\epsilon\cos\theta] +O(\delta^2)$. $\mathsf{Z}$ describes (a
small perturbation of) an ellipse centered at the origin whose great axis is
the segment $[-\delta(1+\mathrm{i}),\delta(1+\mathrm{i})]$ and
whose width is proportional to~$\epsilon$. The corresponding winding number
is $\mathrm{sgn}(x)\mathrm{sgn}(y)$ and therefore the total charge given
by~(\ref{g}) vanishes.

\subsection{An example with a finite but arbitrarily large number of constraints}
\label{subsec:exemple2}
\begin{figure}[!hbt]
\begin{center}
\psfig{file=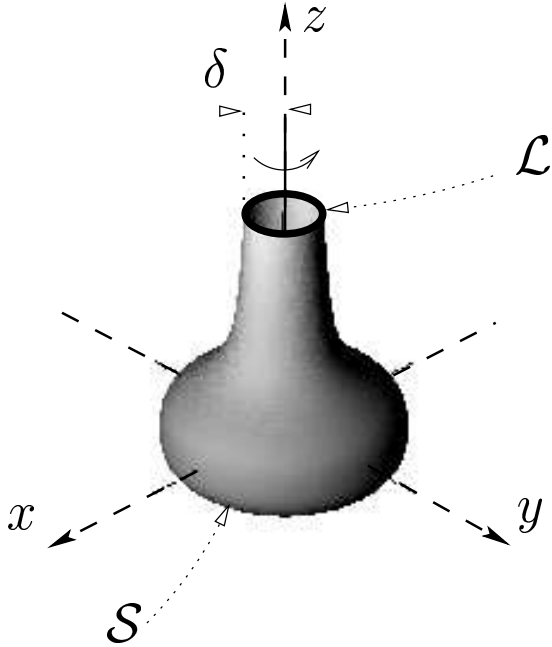}
\caption{    
\label{fig:1Diracstring} The examples considered in section~\ref{subsec:exemple2} 
and~\ref{sec:largecharges} lead to one half Dirac string~$x=y=0$ and~$z\geq0$.}
\end{center}
\end{figure}

Now consider the case
\begin{equation}\label{def:be2}
\text{\bfseries\itshape e}(x,y,z)\ =\
[(xy)^{n},\frac{1}{2}(x^{2n}-y^{2n}),z] +\text{higher order terms}\ ,
\end{equation}
where~$n$ is a strictly positive integer. The propagation of light in a
twisted anisotropic dielectric medium considered in~\cite[see eq. (45)]{Berry86a} 
corresponds to~$n=1$. $\mathcal{D}_+$ is the
half-axis~$x=y=0,z\ge0$. There is only one loop to consider (see
figure~\ref{fig:1Diracstring}).

Let us fix~$z=z_0\ge0$ and~$\delta>0$. Consider the loop
\begin{equation}\label{loop}
         \mathcal{L}\equiv[x=\delta\cos\theta, y=\delta\sin\theta, z=z_0],
         \qquad\theta\in[0,2\pi[\;.
\end{equation}
Then~$\Delta\Phi/(2\pi)$ is the winding number around zero of
\begin{equation}
 \mathsf{Z}(\theta)=\delta^{2n}\left[
 \left(\frac{\sin(2\theta)}{2}\right)^n
 +\frac{\mathrm{i}}{2}\,\bigl(\cos^{2n}\theta
 -\sin^{2n}\theta \bigr) \right] \ .
\end{equation}
If~$n$ is even there is no winding, since the real part of~$\mathsf{Z}$
remains always positive. Therefore~$g=0$. In contrast, if~$n$ is
odd~$\mathsf{Z}(\theta)$ winds clockwise twice, and therefore~$g = -2 g_0$.
Figure~\ref{fig:contacts} illustrates two different contact points
corresponding to the two possible charges.

\section{Arbitrarily large charges}\label{sec:largecharges}

\begin{figure}[!htb]
\begin{center}
\psfig{file=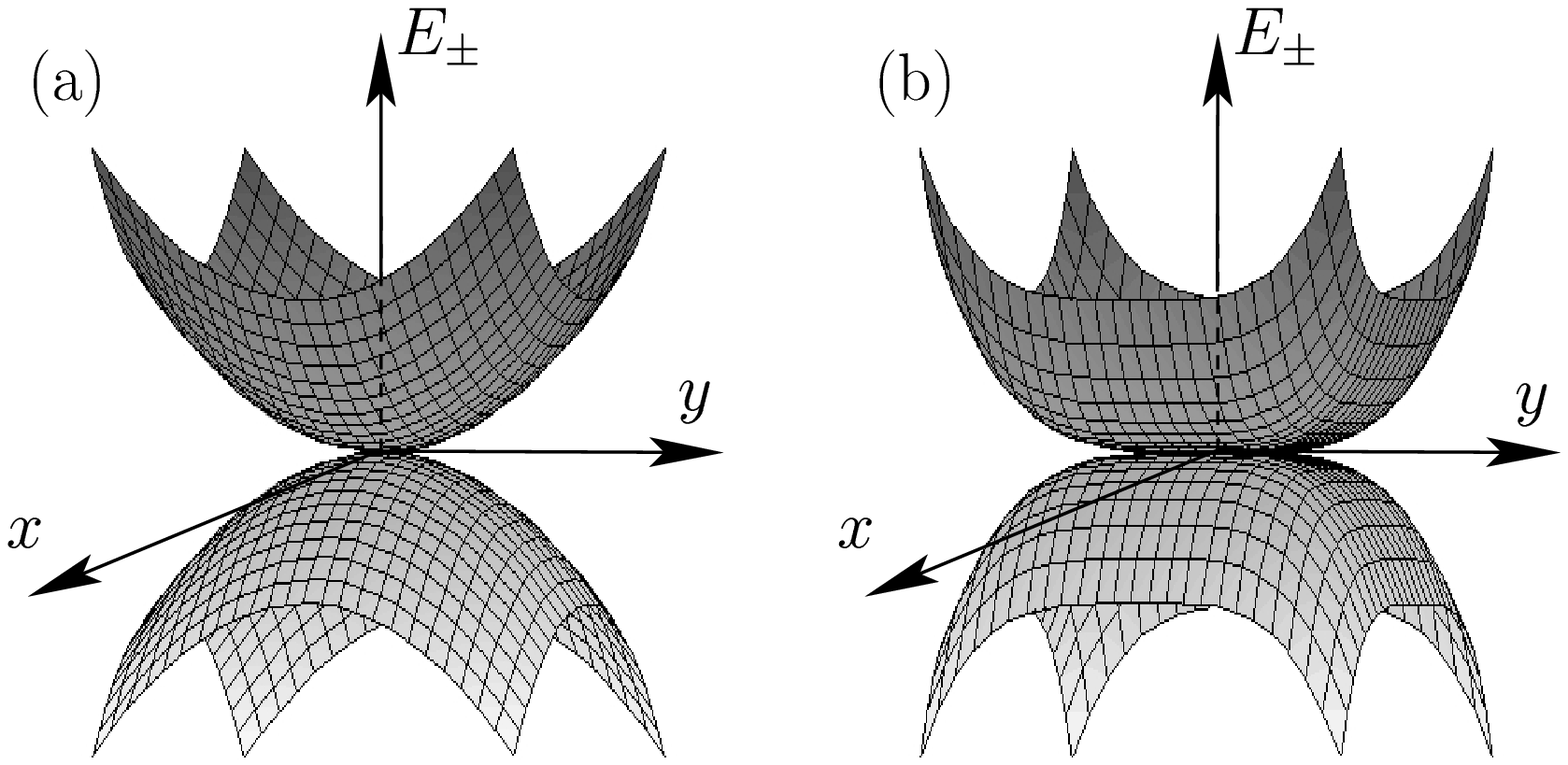,width=\textwidth}
\caption{    
\label{fig:contacts} Geometry of the contact point between energy surfaces for~$z=0$.
Part (a) corresponds to Eq.(\ref{def:be2}) with $n=1$ and to Eq.(\ref{chebye})
with $n=2$. The charge is $g/g_0 = 2$. Part (b) corresponds to
Eq.(\ref{def:be2}) with $n=2$. The charge vanishes. Notice the difference
 in the geometry of the contact between both cases.}
\end{center}
\end{figure}

In order to get arbitrarily large winding numbers for~$\mathsf{Z}$, we make
use of Chebyshev's polynomials, defined by~\cite[chap.V, \S5.7]{Magnus+66a}
\begin{equation} 
\mathsf{C}_n(q)=\cos(n\arccos q) =\frac{1}{2}\left[
\left(q+\mathrm{i}\sqrt{1-q^2}\right)^n
+\left(q-\mathrm{i}\sqrt{1-q^2}\right)^n \right]\;, 
\end{equation} 
where~$n$ is an arbitrary integer and~$q$ a real number satisfying~$-1\leq
q\leq1$. Let us take 
\begin{equation} \label{chebye}
\text{\bfseries\itshape e}(x,y,z)\ =\
\begin{pmatrix}\displaystyle f(\sqrt{x^2+y^2})\ \mathsf{C}_n\!\!
\left[\frac{x}{\sqrt{x^2+y^2}}\right] \\[5ex]\displaystyle g(\sqrt{x^2+y^2})\
\mathsf{C}_n\!\! \left[\cos\left(\frac{\pi}{2n}\right)
\frac{x}{\sqrt{x^2+y^2}} +\sin\left(\frac{\pi}{2n}\right)
\frac{y}{\sqrt{x^2+y^2}} \right] \\[5ex]\displaystyle z \end{pmatrix}
+\text{higher order terms}\;.
\end{equation} 
$f(q)$ and $g(q)$ are some regularizing real functions that vanish at $q=0$
faster than any singular denominator coming from~$\mathsf{C}_n$. An
appropriate choice is, for instance, $f(q)=g(q)=q^n$.
Though~$\mathsf{C}_n$ vanishes~$n$ times in~$[-1,1]$, one can always choose a
small neighborhood where~$\mathbf{r}=\mathbf{0}$ is the unique point
in~$\mathcal{M}$. As in the last example, the set~$\mathcal{D}_+$ is made of
one Dirac half string on the half-axis~$x=y=0,z\ge0$. Constructing the same
loop as before one gets
\begin{equation}
\mathsf{Z}(\theta)=f(\delta)\cos(n\theta)+\mathrm{i}\,g(\delta)\sin(n\theta) \ ,
\end{equation} 
which corresponds to~$g = ng_0$. This provides an explicit example of a
monopole carrying an arbitrarily large charge, and therefore creating
arbitrarily large phase shifts for a loop located close to the degeneracy in
parameter space. The contact between energy surfaces with $n=2$ is illustrated
in part (a) of figure \ref{fig:contacts}, which is identical to the geometry
of the contact defined by Eq.(\ref{def:be2}) with $n=1$ (up to an irrelevant
scaling factor).

\section{Discussion}\label{sec:discussion}

Spectral degeneracies with charges higher than one have been observed in
several contexts. For instance, via the variations of Chern numbers in quantum
systems with mixed and chaotic classical dynamics
\cite{Leboeuf+90a,Leboeuf+92a}. These integer topological invariants measure
the transport (Hall conductivity) of each band of a doubly periodic system.
Values of $g/g_0 = 2$ and $3$ were found (cf in Fig.~2 of
Ref.~\cite{Leboeuf+90a} the contact points at $\gamma = 0.7106$ between the
second and third levels, and at $\gamma = 0.4477$ between the fourth and fifth
levels, respectively). A closer analysis of the geometry of the contact points
reveals that these two degeneracies coincide with the parabolic and cubic
contacts described by Eq.(\ref{chebye}) with $n=2$ and $3$, respectively. For
instance, the $g/g_0 = 2$ contact is parabolic in the Bloch angles and linear
in the parameter $\gamma$.

We have also exhibited the possibility of having a contact point with total
charge equal to zero. Although its presence will in general be difficult to
detect, it would be interesting to find an explicit quantum mechanical model
displaying this curious feature.

In the electromagnetic analogy based on the general formula, the ``sources''
of the geometric phase in parameter space are the spectral degeneracies plus
some additional currents not related to them. In the particular (and generic)
case of a diabolical contact, the contribution of the degeneracy is simple: it
acts as a pure monopole charge, thus contributing to $\mathrm{div}\,
\mathbf{B}$ but not to $\mathbf{curl}\, \mathbf{B}$. However, for an arbitrary
contact point -- defined by Eq.(\ref{eq:B}) -- the situation is different. On
the one hand, the divergence of $\mathbf{B}$ is not simply determined by a
charge, but more generally by a {\sl charge distribution} (cf
Eq.(\ref{eq:divexpansion})). The latter can contain higher multipole moments.
Moreover, the $\mathrm{curl}$ of $\mathbf{B}$ need not be zero. In the present
study, we have only considered a particular aspect of the sources of geometric
phase associated to the field produced by a degeneracy, namely the total
charge of the distribution. A more complete classification scheme is clearly
needed.

\end{document}